\documentclass[sigconf]{acmart}

\renewcommand\footnotetextcopyrightpermission[1]{} % removes footnote with conference information in first column

\setcopyright{none}

\acmConference[SBES 2026]{40th Brazilian Symposium on Software Engineering}{September 8–12, 2026}{São Paulo, SP, Brazil}

\AtBeginDocument{
    
}

\begin{document}

%% The "title" command has an optional parameter,
%% allowing the author to define a "short title" to be used on the page 
%% headers.
\title{SoftBoard: A Multi-Agent Tool for the Creation and Evaluation of Low-Fidelity Prototypes}

%% The "author" command and its associated commands are used to define
%% the authors and their affiliations.
%% Of note is the shared affiliation of the first two authors, and the
%% "authornote" and "authornotemark" commands
%% used to denote shared contribution to the research.
\author{Gabriel R. S. Scapim}
\affiliation{
  \institution{State University of Maringá}
  \city{Maringá}
  \country{Brazil}
}
\email{scapimgabriel@gmail.com}

\author{Gislaine C. L. Leal}
\affiliation{
  \institution{State University of Maringá}
  \city{Maringá}
  \country{Brazil}
}
\email{gclleal@uem.br}

\author{Guilherme C. Guerino}
\affiliation{
  \institution{State University of Paraná}
  \city{Apucarana}
  \country{Brazil}
}
\email{guilherme.guerino@ies.unespar.edu.br}

%% By default, the full list of authors will be used on the page
%% headers. This list is often too long and will overlap
%% other information printed in the page headers. 
%% This command allows the author to define a more concise list
%% of authors' names for this purpose.
\renewcommand{\shortauthors}{Scapim et al.}
\renewcommand{\shorttitle}{SoftBoard: A Multi-Agent Tool for the Creation and Evaluation of Low-Fidelity Prototypes}

%% The abstract is a short summary of the work the paper presents.
\begin{abstract}
User Experience (UX) is recognized as a critical factor for the success of digital products, particularly in software startups, environments marked by time constraints, limited resources, and low maturity in design practices. Building Minimum Viable Products (MVPs) through low-fidelity prototyping represents a well-established strategy for rapid validation cycles at reduced cost. A systematic literature mapping, however, revealed gaps in the ecosystem of available tools: a predominance of general-purpose solutions adapted for prototyping, the absence of integrated methodological guidance, and the incipient use of Artificial Intelligence in the design process. This paper presents SoftBoard, a web-based tool for the creation and evaluation of low-fidelity prototypes in the context of MVP development. The tool integrates a prototype editor, team-based project organization, and a multi-agent system based on large language models that supports requirements elicitation and refinement, automates prototype generation, and evaluates interface quality based on usability heuristics. This integration aims to reduce reliance on prior UX expertise, standardize the prototyping process, and support teams in building MVPs aligned with user needs. As future work, a feasibility study with software professionals is currently underway.\newline
\textbf{Demo video: https://doi.org/10.5281/zenodo.20369601}
\end{abstract}

%% Keywords. The author(s) should pick words that accurately describe
%% the presented work. Separate the keywords with commas.

%% This command processes the author, affiliation, and title
%% information and builds the first part of the formatted document.
%% "frenchspacing" avoids an additional space after a period at the end of a sentence.
\frenchspacing
\maketitle

\section{Introduction}
User Experience (UX) plays a central role in the development of digital products, encompassing users’ perceptions and responses resulting from the use or anticipated use of a system \cite{ISO9241-210:2019}. Ensuring an adequate UX from the early stages of development is particularly critical in Software Startups, environments characterized by strong time pressure, limited resources, small teams, and high market uncertainty \cite{6898758}. In such contexts, design decisions are often made empirically and with limited methodological support, which may lead to products that are misaligned with user needs, increased rework, and low user adoption \cite{Hokkanen2016, MaturroEtAl2021}.

One strategy adopted to mitigate these risks is the development of a Minimum Viable Product (MVP), which enables short and iterative cycles of validated learning \cite{ries2011startup}. To fulfill this role, an MVP must balance simplicity with the ability to generate meaningful evidence about user behavior. In this context, low-fidelity prototypes, such as wireframes and digital sketches, are particularly suitable for the early stages of development, as they enable rapid iterations, low modification costs, and a focus on interface structure and navigation flow rather than visual aesthetics \cite{9582602, brooks2023smell}. Nevertheless, many Software Startups still face difficulties in structuring MVPs that achieve this balance, whether due to a lack of methodological knowledge or the absence of tools and processes tailored to their context \cite{MaturroEtAl2021, Nguyen-Duc2017}.

To systematize the development of UX-oriented MVPs in Software Startups, the StartFlow method was proposed \cite{guerino2026arXiv}. The method adopts wireflows as its central artifact, visual representations that combine wireframes and user flow maps, enabling the simultaneous visualization of screen structures and navigation transitions. The process is organized into three main stages: identifying and structuring MVP functionalities, transforming these functionalities into wireflows, and reviewing the developed prototype. Feasibility studies with software professionals indicate that StartFlow contributes to prioritization, reduces inconsistencies, facilitates alignment among product, technology, and UX teams, and promotes a more strategic prototyping process \cite{guerino2026arXiv}.

Despite these contributions, the application of StartFlow relies exclusively on textual documentation, leaving the choice of prototyping tools to practitioners. The absence of a dedicated tool may hinder the adoption of the method, increase variability in outcomes, and make its application highly dependent on the experience and preferences of individual team members. To investigate this scenario, a Systematic Mapping Study was conducted on low-fidelity prototyping tools. The results revealed gaps, including functional limitations in existing tools, limited incorporation of structured design methods and heuristics, and the still incipient adoption of Artificial Intelligence (AI) to support the prototyping process \cite{scapim:2025}.

In response to these challenges, this paper presents SoftBoard, a web-based tool designed to operationalize the StartFlow method for the creation and evaluation of low-fidelity prototypes. SoftBoard integrates a wireframe and wireflow editor, team-based project management, and a guided workflow (wizard) structured around the three stages of StartFlow, supported by a multi-agent system based on Large Language Models (LLMs). Within this system, specialized agents collaborate throughout the process by supporting requirements elicitation and validation, automating wireflow generation, and evaluating interfaces based on StartFlow-specific heuristics as well as Nielsen’s Usability Heuristics \cite{nielsen1994heuristic}. In this way, SoftBoard aims to lower the adoption barrier of the StartFlow method, standardize the prototyping process, and support teams with limited UX expertise in developing MVPs that are aligned with user needs.

The remainder of this paper is organized as follows. Section 2 presents the theoretical background and related work. Section 3 describes SoftBoard, its architecture, and its main functionalities. Section 4 presents a usage example. Finally, Section 5 concludes the paper.

\section{Background}
This section presents the theoretical background that underpins this work. First, concepts related to low-fidelity prototyping and MVP development are discussed. Next, the StartFlow method is described. Finally, related tools and studies on low-fidelity prototyping are presented.

\subsection{MVPs and Low-Fidelity Prototyping}

The Minimum Viable Product (MVP) is a central concept in digital product development and is defined as an initial version of a product designed to maximize validated learning about customers with the least possible effort \cite{ries2011startup, MaturroEtAl2021}. Rooted in the Lean Startup movement, the MVP emphasizes the delivery of a minimal functional solution that generates real value while enabling the rapid collection of user feedback \cite{ARDITO2014542}. Literature reviews indicate that concepts such as a minimum set of features, an initial prototype, reduced effort, and customer feedback are recurrent elements in MVP definitions \cite{DavideEtAl2016}.

The development of MVPs in startups takes place under constraints related to time, team size, and financial resources, often accompanied by communication difficulties and product--market misalignment \cite{MaturroEtAl2021}. Recent studies reinforce that the MVP should be understood as an experimental mechanism for value validation rather than merely a reduced technical artifact \cite{Javdani_Gandomani2024-za}. In this context, the MVP also plays a guiding role by supporting the prioritization of essential and testable functionalities, particularly in teams with lower levels of technical maturity \cite{Guerino_Martinelli_Choma_Leal_Balancieri_Zaina_2024}. The adoption of more structured practices and processes contributes to improving both the speed and quality of decision-making in environments characterized by uncertainty \cite{Nguyen-Duc2017}.

Interface prototyping plays a central role in this process, particularly within the User-Centered Design (UCD) approach, which prioritizes users' needs and limitations throughout the development of digital solutions \cite{abras2004user,GOMES2024}. The iterative creation and evaluation of prototypes enable teams to validate ideas, anticipate problems, and establish early visual representations before implementation \cite{mccurdy2006breaking,Romero2024}. Low-fidelity prototypes are characterized by their simplicity, low cost, and ease of modification, making them suitable for exploring initial concepts and structuring interaction flows without requiring advanced visual or technical details \cite{10.1007/978-3-540-74796-3_16, 10.1145/3411764.3445520}. Examples include paper sketches, wireframes, and rapidly created digital representations \cite{brooks2023smell,10.1145/3411764.3445520}, which are considered particularly appropriate for MVP development due to their support for rapid iteration and idea validation \cite{9582602}.

\subsection{The StartFlow Method}

StartFlow was designed to support startups during the early stages of digital product conception, enabling teams to represent the essential functionality of an MVP in a clear and cost-effective manner before implementation through the use of wireflows \cite{guerino2026arXiv}. These artifacts combine screen sketches and navigation flows, allowing teams to visualize the expected user behavior during interaction with the system.

The method organizes the work into three main stages: identifying and structuring functionalities, transforming these functionalities into wireflows, and reviewing the interaction flows. This process guides teams in understanding how each operation will be performed by users, which interface elements are required at each stage, and how transitions between screens should occur. As a result, the approach promotes alignment among technology, product, and UX teams while making the rationale behind design decisions more transparent and collaborative \cite{guerino2026arXiv}.

The wireflows produced during the method allow the simultaneous visualization of the minimal interface structure and user navigation paths. Elements such as buttons, input fields, and menus are represented in a simplified manner, while connectors indicate each step of the interaction flow. This representation enables teams to examine alternative scenarios, identify dependencies among functionalities, and anticipate error situations or unlikely navigation paths \cite{guerino2026arXiv}.

Feasibility studies conducted with software professionals indicate that the method's guiding questions serve as a reasoning framework, helping participants organize priorities, refine requirements, and reduce common omissions during the early stages of product conception. Participants also highlighted that the visual representation of flows improves the collective understanding of product behavior, facilitates communication among teams, and can be applied at a low cost, even by professionals without formal UX training \cite{guerino2026arXiv}.

\subsection{Related Tools}

To identify the state of the art in low-fidelity prototyping tools, a Systematic Mapping Study was conducted \cite{scapim:2025}. The initial search returned 1,493 publications, and after applying the inclusion and exclusion criteria, 37 primary studies were selected, from which 22 tools were identified. Among the most frequently cited tools were Figma\footnote{https://www.figma.com}, Balsamiq\footnote{https://balsamiq.com}, and Microsoft PowerPoint\footnote{https://www.microsoft.com/pt-br/microsoft-365/powerpoint}, followed by Adobe XD\footnote{https://adobexdplatform.com}, Pencil Project\footnote{https://pencil.evolus.vn}, and Sketch\footnote{https://www.sketch.com}.

One of the main findings concerns the predominance of general-purpose tools adapted for prototyping. Solutions such as Figma, PowerPoint, Adobe XD, and Sketch were not originally designed specifically for low-fidelity prototyping, yet they account for most of the citations in the analyzed studies. This pattern suggests that practitioners rely on these tools due to familiarity, integration with established workflows, or the lack of sufficiently comprehensive dedicated alternatives. Although these tools provide a wider range of features, their use may reduce methodological guidance and focus during the early stages of design.

The study also revealed that only a small portion of the analyzed tools incorporate AI-based functionalities, such as automatic prototype generation or suggestions for layout and usability improvements. Uizard\footnote{https://uizard.io} stands out by converting hand-drawn wireframes into digital prototypes and providing automated suggestions. Nevertheless, this approach remains relatively uncommon among the available tools.

Another finding was the limited integration of structured methods, heuristics, and design guidelines within prototyping tools. Only 5 of the 22 identified tools incorporate some form of method or guideline, and only 2 are dedicated exclusively to low-fidelity prototyping. Among the existing examples, Figma and Adobe XD allow the integration of guidelines through plugins, while WOZ Pro incorporates the Wizard of Oz method for evaluation \cite{10.1145/1240866.1241023}. However, such integration is indirect and depends on the initiative of practitioners. In most tools, there is no built-in guidance on how to structure interface decisions, leaving users responsible for ensuring the quality and consistency of the resulting solutions.

Recent studies point to a growing trend in the use of Artificial Intelligence to support activities related to design and requirements engineering. A conceptual framework for generating user stories from real-time conversations with users was proposed to reduce dependence on pre-existing requirements documentation \cite{sbes}. Similarly, a ChatGPT-based tool was developed to support the automated identification and classification of usability defects according to Nielsen’s heuristics, assisting teams that lack dedicated usability specialists \cite{SolariaGPT}.

\section{SoftBoard}
This section presents SoftBoard, describing its overview, target users, requirements, system architecture, and main implemented functionalities. For improved readability and visualization, high-resolution versions of both architectural diagrams are available in the artifacts section of the replication package, accessible at \url{https://doi.org/10.5281/zenodo.20369601}.

\subsection{Overview and Target Users}

SoftBoard is a web-based tool designed to support the creation and evaluation of low-fidelity prototypes. Its primary goal is to lower the adoption barrier by providing a structured environment that guides practitioners from MVP functionality definition to wireflow generation and evaluation, without requiring prior experience in UX or design tools.

The tool is primarily intended for professionals working in software startups, 
who need to build MVPs quickly and systematically but often lack dedicated UX 
specialists within their teams.

SoftBoard is proprietary software, all rights reserved. The source code is not 
publicly available at this time, as a feasibility study with software professionals 
is currently underway.

\subsection{Requirements}

The following requirements were defined for SoftBoard based on the findings of the Systematic Mapping Study \cite{scapim:2025} and the operational needs of the StartFlow method \cite{guerino2026arXiv}. The requirements were organized into three modules: Platform, Editor, and StartFlow.

\subsubsection{Platform}

\begin{itemize}
\item \textbf{FR1. User registration:} The tool shall allow users to register and edit their personal information.

\item \textbf{FR2. Authentication:} The tool shall provide secure authentication, ensuring that each user can access only their own teams and information.

\item \textbf{FR3. Password recovery:} The tool shall allow users to recover access through password reset functionality.

\item \textbf{FR4. Team management:} The tool shall allow users to create, edit, and delete teams, which serve as collaborative workspaces.

\item \textbf{FR5. Member management:} The tool shall allow users to add and remove team members, with role-based permission control.

\item \textbf{FR6. Board management:} The tool shall allow users to create, edit, list, and delete boards within a team. Each board represents a prototyping project and centralizes the editor, requirements, interaction history, and the current StartFlow stage.

\item \textbf{FR7. Board sharing:} The tool shall allow users to generate a public link for the interactive visualization of completed boards, facilitating communication with stakeholders.
\end{itemize}

\subsubsection{Editor}

\begin{itemize}
\item \textbf{FR8. Component gallery:} The tool shall provide a gallery of essential components for mobile interface prototyping, including buttons, input fields, icons, text elements, shapes, and toggles, all configurable through visual properties.

\item \textbf{FR9. Drag-and-drop support:} The tool shall allow users to add and position components on the canvas through drag-and-drop interactions, with support for hierarchical relationships between components and screens.

\item \textbf{FR10. Component editing:} The tool shall allow users to resize, reposition, and reorder components across layers, providing automatic alignment guides during manipulation.

\item \textbf{FR11. Wireflow creation:} The tool shall allow users to create multiple screens and establish navigation connections between them through linked components, representing the MVP interaction flows.

\item \textbf{FR12. Interactive mode:} The tool shall provide an interactive mode that enables navigation across connected screens, simulating application behavior directly within the editor.
\end{itemize}

\subsubsection{StartFlow}

\begin{itemize}
\item \textbf{FR13. Guided workflow:} The tool shall structure each board into three sequential stages aligned with StartFlow: requirements organization, wireflow construction, and review, with controlled progression between stages.

\item \textbf{FR14. Requirements management:} The tool shall allow users to create, edit, and delete requirements in the form of user stories.

\item \textbf{FR15. Conversational agent:} The tool shall provide a conversational agent that assists practitioners throughout the wizard stages, adapting its behavior, tools, and instructions according to the current stage.

\item \textbf{FR16. Automatic wireflow generation:} The tool shall allow users to trigger automatic wireflow generation from the defined requirements through a multi-agent system executed asynchronously, with notification upon completion.

\item \textbf{FR17. Automatic UX evaluation:} The tool shall automatically evaluate generated wireflows based on UX heuristics, assign Likert-scale scores ranging from 1 to 5, and provide improvement suggestions for each criterion.
\end{itemize}

\subsection{Architecture}

\begin{figure}[ht]
  \centering
  \includegraphics[width=0.4\linewidth]{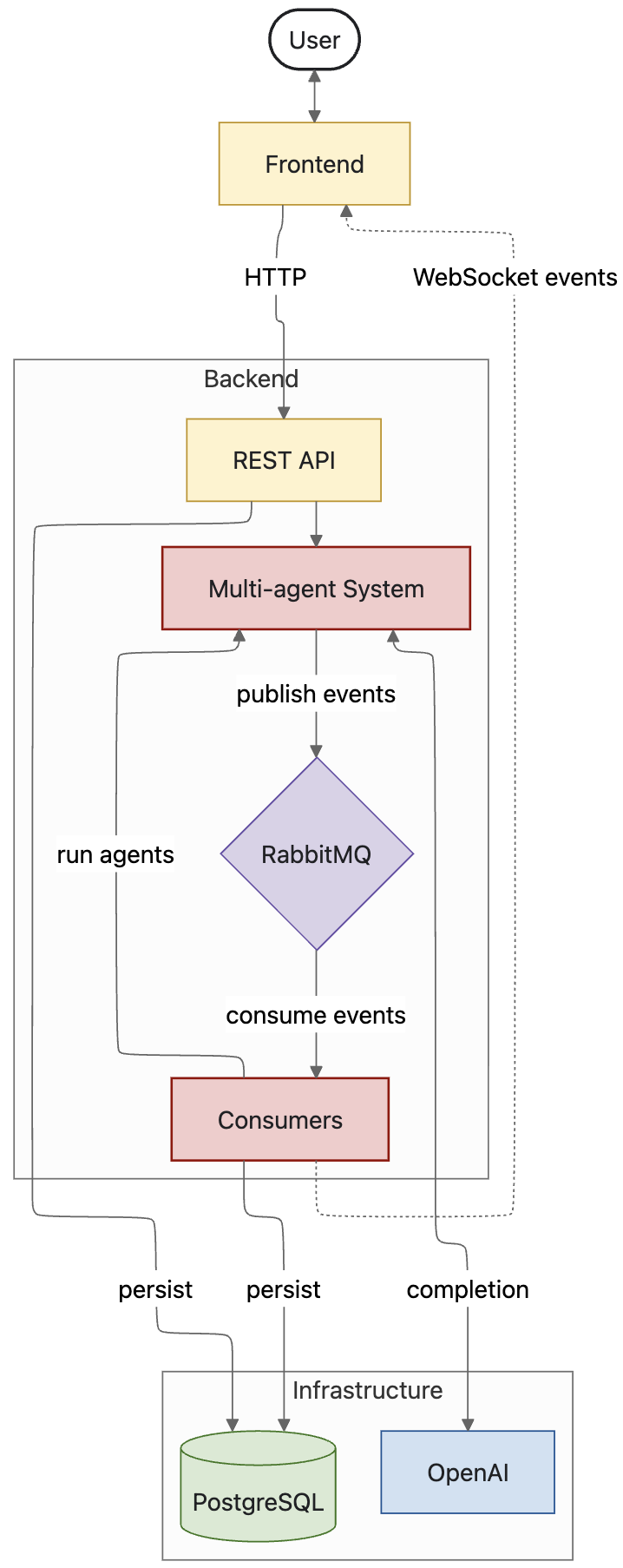}
  \caption{SoftBoard architecture.}
  \label{fig:arquitetura}
  \Description{A block diagram showing the system architecture of SoftBoard divided into Frontend, Backend, and Infrastructure. The User interacts bidirectionally with the Frontend, which communicates with the Backend via HTTP requests and receives async WebSocket events back. Inside the Backend box, requests flow from a REST API to a Multi-agent System, which publishes events to a RabbitMQ message broker. The broker distributes messages to Consumers that run agents, communicating back to the Multi-agent System. Both the REST API and Consumers persist data into a PostgreSQL database in the Infrastructure layer, while the Multi-agent System connects to OpenAI for LLM completions.}
\end{figure}

SoftBoard is a web application organized into three layers: frontend, backend, and infrastructure, as illustrated in Figure~\ref{fig:arquitetura}. The entire application was developed using the TypeScript programming language\footnote{https://www.typescriptlang.org}.

\subsubsection{Frontend}

The frontend was implemented using React\footnote{https://react.dev}. The interface includes the platform screens (authentication, profile, teams, and members) as well as the workspace environment, referred to as \textit{boards}. A board serves as the central integration point of the user experience, bringing together the wireframe and wireflow editor, chat interface, requirements management, and the wizard that guides practitioners through the stages of the method.

Communication with the backend occurs through HTTP requests for synchronous operations and WebSockets for real-time notifications, such as wireflow generation completion and UX evaluation delivery. The interface is available in both Portuguese and English.

\subsubsection{Backend}

The backend was developed using Node.js\footnote{https://nodejs.org} and Express\footnote{https://expressjs.com}. The REST API handles authentication, password recovery, team and permission management, as well as board and requirement administration. During the wizard execution, the conversational agent operates synchronously through API requests.

When more computationally intensive tasks are required, an event is published to RabbitMQ\footnote{https://www.rabbitmq.com} and processed asynchronously by consumers responsible for executing specialized agents. Once processing is completed, the frontend is notified via WebSocket.

\subsubsection{Infrastructure}

PostgreSQL\footnote{https://www.postgresql.org} stores all application data, including users, teams, boards, requirements, wireflows, chat messages, and UX evaluations. RabbitMQ decouples computationally intensive processing from the HTTP request cycle. The entire infrastructure is containerized using Docker Compose\footnote{https://docs.docker.com/compose}.

The language models used by the multi-agent system are accessed through the OpenAI API\footnote{https://openai.com}.

\subsection{Multi-Agent System}

\begin{figure}[ht]
  \centering
  \includegraphics[width=1\linewidth]{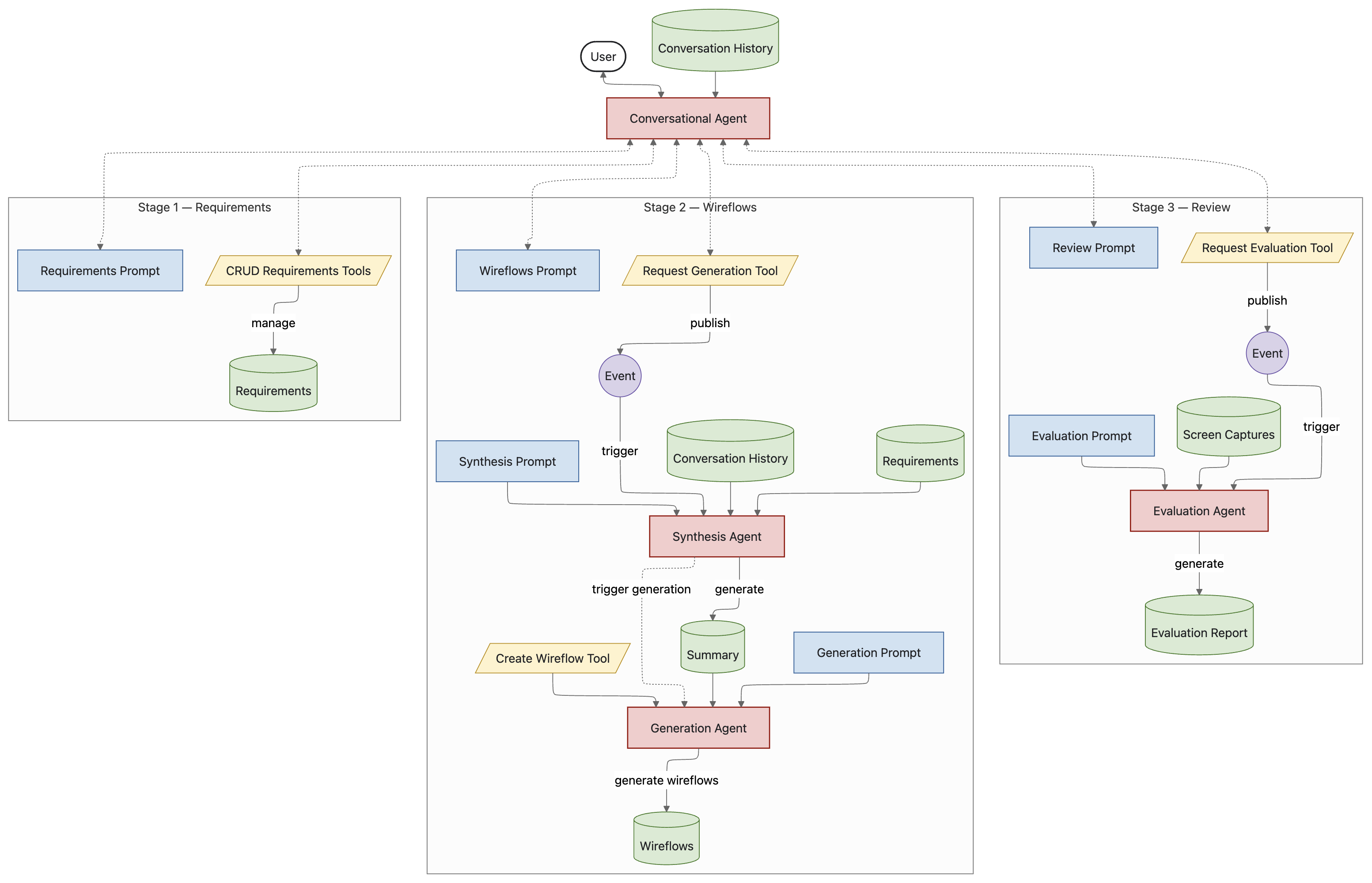}
  \caption{SoftBoard multi-agent system.}
  \label{fig:multiagente}
    \Description{A detailed workflow diagram of the SoftBoard multi-agent system interaction across three stages. At the top, a User and a Conversation History database connect to a central Conversational Agent box. Below, the process splits into three large boxes: Stage 1 Requirements, containing a Requirements Prompt and CRUD tools managing a Requirements database; Stage 2 Wireflows, where a Wireflows Prompt publishes a request generation event that triggers a Synthesis Agent using conversation history and requirements to output a Summary, which then triggers a Generation Agent to create wireflows; and Stage 3 Review, where a review request triggers an Evaluation Agent using review prompts and screen captures to generate a final Evaluation Report.}
\end{figure}

The integration with the StartFlow method is achieved through four specialized agents organized into three stages that mirror the wizard workflow (Figure~\ref{fig:multiagente}). All agents share the same underlying abstraction, differing primarily in the LLM model employed, the available tools, and the instructions provided, which are grounded in the StartFlow guiding questions. The conversational agent uses GPT-4o, whereas the synthesis, generation, and evaluation agents use GPT-5.5.

During Stage 1 (Requirements), the conversational agent supports the identification and structuring of MVP functionalities as user stories, encouraging reflection on essential user needs and behaviors. Throughout the interaction, the agent incrementally creates, edits, and organizes requirements.

During Stage 2 (Wireflows), when generation is requested, the conversation history and consolidated requirements are submitted for asynchronous processing. First, the synthesis agent verifies the sufficiency of the available information and produces a structured MVP summary. Next, the generation agent transforms this summary into wireflows composed of interconnected screens, which are then persisted in the database. If the information is deemed insufficient, the system returns an error message to the user.

During Stage 3 (Review), images of the generated wireflows are analyzed by an evaluation agent with computer vision capabilities. The system examines eight criteria: four derived from StartFlow and four based on Nielsen’s Usability Heuristics \cite{nielsen1994heuristic}. Each criterion receives a score ranging from 1 to 5, accompanied by suggestions for improvement.

\section{Usage example}
This section presents an illustrative usage scenario of SoftBoard to demonstrate the tool's operation in practice. The example simulates the conception of an MVP for a food delivery application.

\begin{figure}[ht]
\centering
\includegraphics[width=0.95\linewidth]{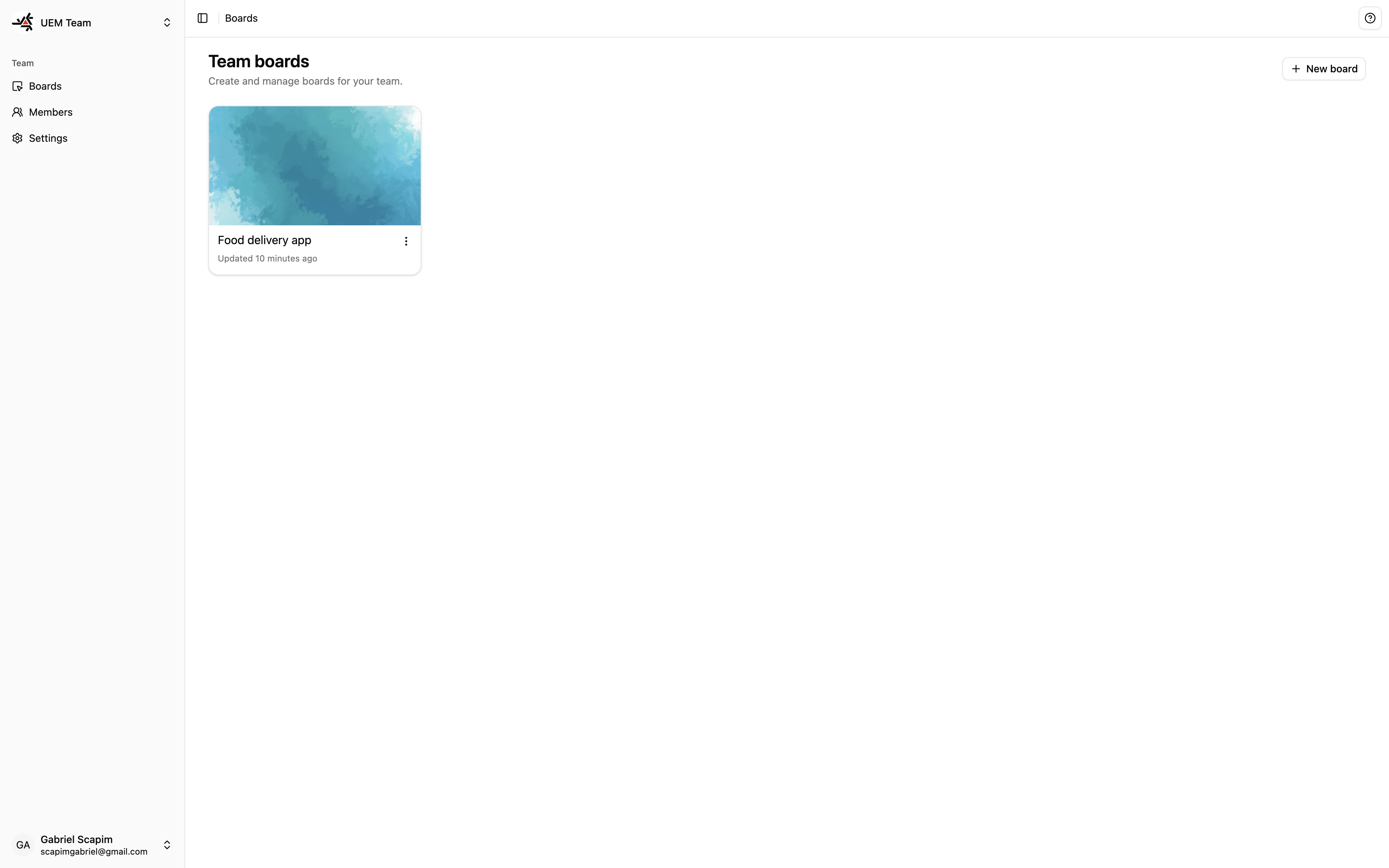}
\caption{SoftBoard board management screen.}
\Description{SoftBoard web interface dashboard showing the "Team boards" section for "UEM Team". On the left side, a vertical navigation menu displays options for Boards, Members, and Settings. The main content area shows a single project card titled "Food Delivery App" with a blue watercolor-style thumbnail preview. A "New board" button is located in the upper right corner.}
\label{fig:boards}
\end{figure}

The workflow begins on the platform's board management screen, where projects are organized by teams. Figure~\ref{fig:boards} shows the main dashboard, in which the board for the ``Food Delivery App'' project can be viewed.

\begin{figure}[ht]
\centering
\includegraphics[width=0.95\linewidth]{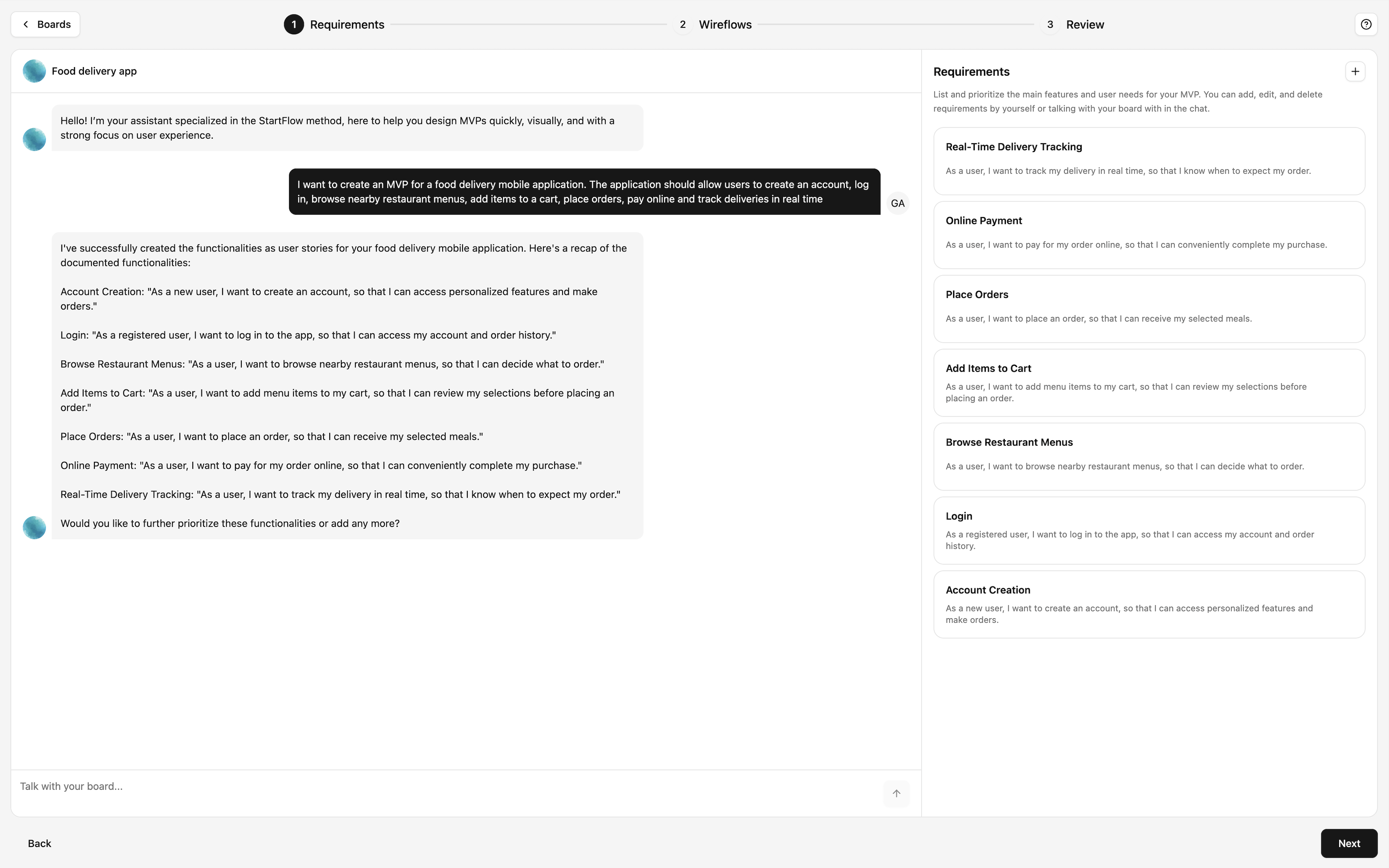}
\caption{Interaction with the conversational agent and requirements structuring.}
\Description{The requirements wizard interface divided into two main panels under a top progress bar highlighting the "Requirements" step. The left panel contains a chat log with an AI assistant discussing user stories for a food delivery MVP. The right panel displays a consolidated vertical list of card components representing specific requirements like "Real-Time Delivery Tracking", "Online Payment", and "Place Orders".}
\label{fig:requirements}
\end{figure}

Upon accessing the board, the user is directed to the wizard responsible for identifying and organizing MVP functionalities. At this stage, the user interacts with a conversational agent that employs StartFlow's guiding questions to support requirements elicitation and refinement. Figure~\ref{fig:requirements} presents the result of this interaction: the conversation history is displayed on the left, while the consolidated list of requirements is shown on the right.

Progression through the workflow is controlled by the tool: users cannot proceed to the next stage unless at least one requirement has been defined, ensuring that subsequent activities are grounded on a consistent minimum scope. This activity illustrates how SoftBoard transforms a typically informal task into a guided and structured process, reducing dependence on prior UX expertise and promoting greater consistency in MVP scope definition.

\begin{figure}[ht]
\centering
\includegraphics[width=0.95\linewidth]{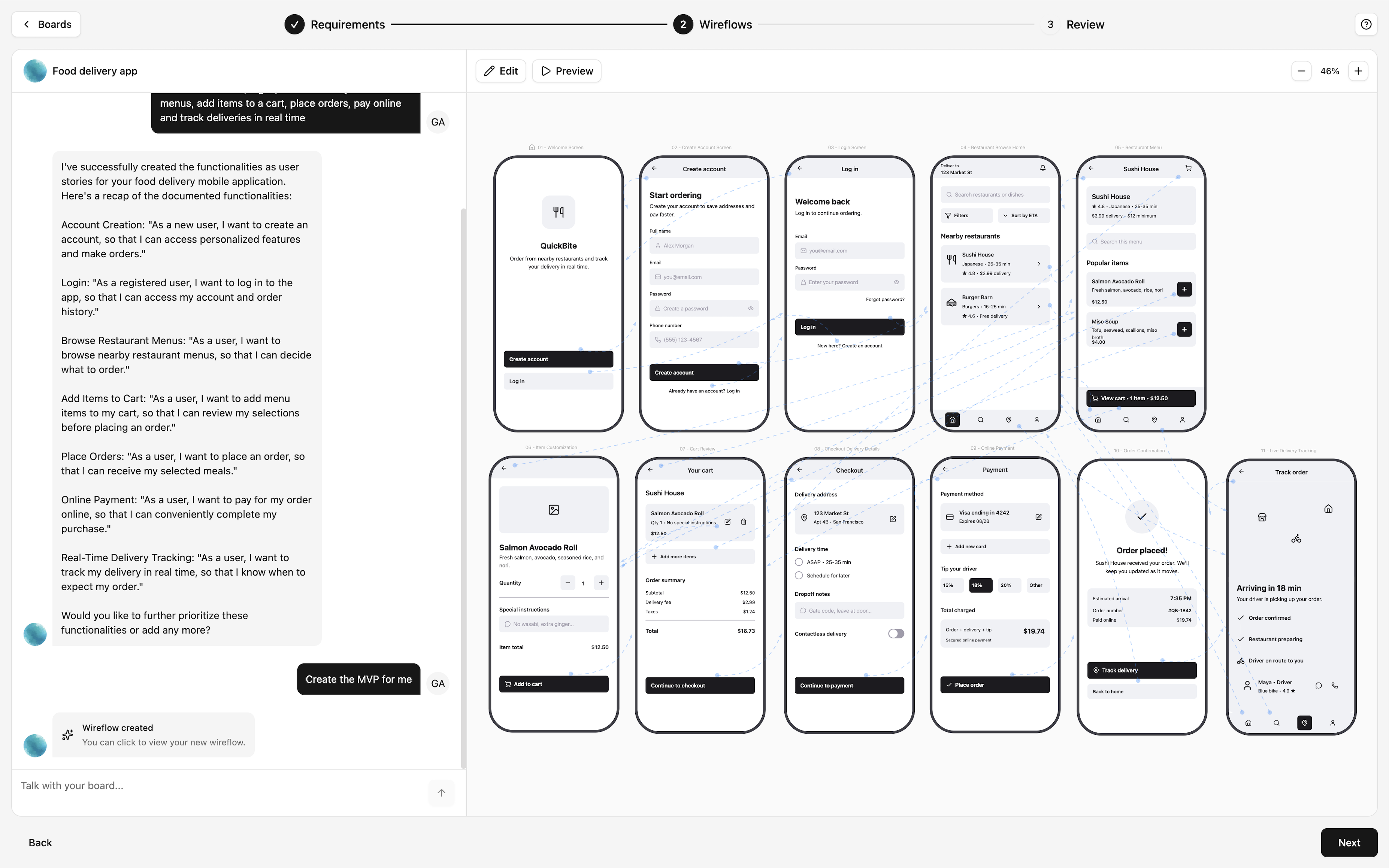}
\caption{Request for automatic wireflow generation.}
\Description{The wireflow generation step interface under a top progress bar highlighting the "Wireflows" step. On the left, the chat window shows a user request saying "Create the MVP for me" and a confirmation notification reading "Wireflow created". The main central canvas displays an interconnected web of low-fidelity mobile wireframe screens linked together by dashed blue user flow lines.}
\label{fig:wireflows_chat}
\end{figure}

Once the requirements have been defined, the user proceeds to the wireflow construction stage. At this point, the automatic generation agent can be triggered to create the initial screen structure and navigation connections based on the previously identified functionalities. Figure~\ref{fig:wireflows_chat} shows a request for prototype generation and the beginning of the asynchronous processing.

\begin{figure}[ht]
\centering
\includegraphics[width=0.95\linewidth]{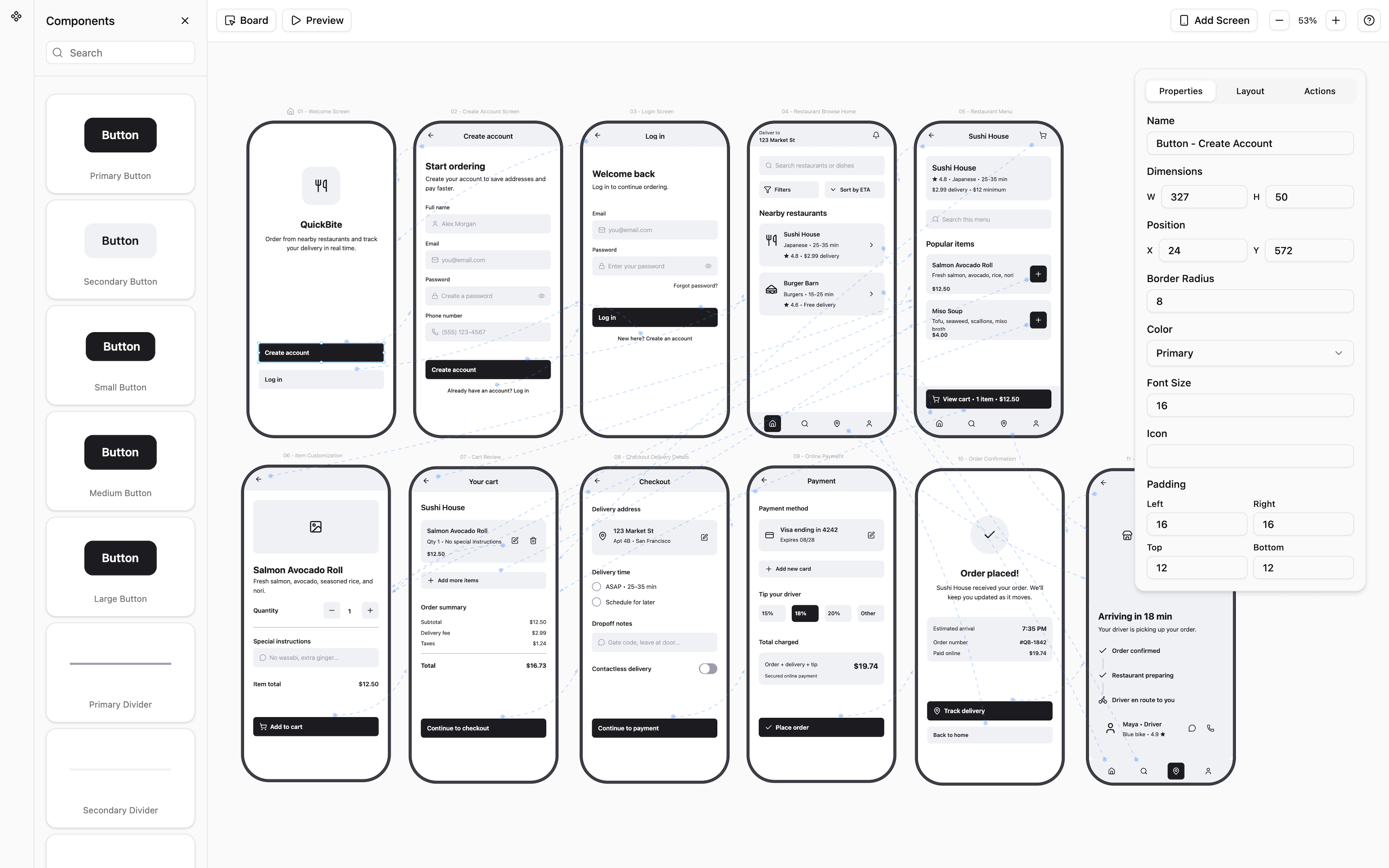}
\caption{Wireflow editor with screen flow and properties panel.}
\Description{The wireflow editing interface. The left sidebar displays a "Components" library containing reusable low-fidelity elements like buttons and dividers. The central canvas shows twelve sequential mobile app screens mapped out with UI flows. The right sidebar shows the "Properties" panel detailing the dimensions, positioning, color, and padding configuration of a selected "Create Account" button.}
\label{fig:editor_detail}
\end{figure}

After generation, the wireflows remain fully editable within the integrated editor. Figure~\ref{fig:editor_detail} presents the editing environment in use, highlighting the canvas containing the complete screen flow, the low-fidelity component gallery, and the properties panel of the selected element.

As in the previous stage, navigation is constrained: the tool requires the existence of at least one screen containing at least one component before allowing progression to the review stage, ensuring that the evaluation is performed on a minimally functional artifact.

\begin{figure}[ht]
\centering
\includegraphics[width=0.95\linewidth]{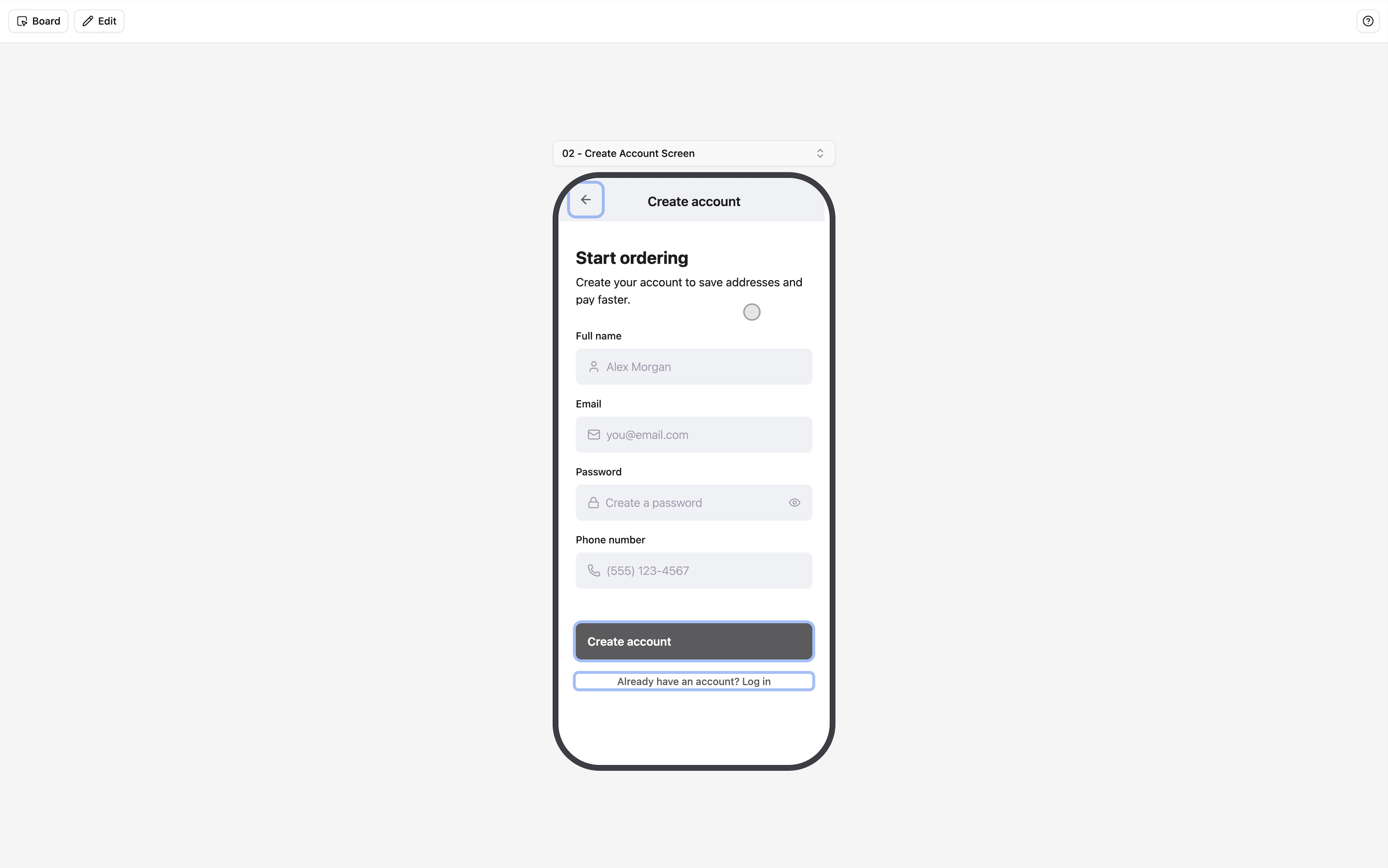}
\caption{Interactive mode for navigation between connected screens.}
\Description{The interactive preview mode inside the editor focusing on a single, enlarged low-fidelity mobile screen titled "02 - Create Account Screen". The interface simulates the registration form layout with text fields for Full name, Email, Password, and Phone number, alongside a primary "Create account" button and a back arrow navigation shortcut.}
\label{fig:interactive_mode}
\end{figure}

In addition to static editing, the editor provides an interactive mode that enables navigation between connected screens, simulating application behavior directly within the editing environment. This functionality allows users to traverse the defined navigation flows, validate the consistency of screen transitions, and identify gaps in the user journey. Figure~\ref{fig:interactive_mode} presents an example of this navigation mode.

\begin{figure}[ht]
\centering
\includegraphics[width=0.95\linewidth]{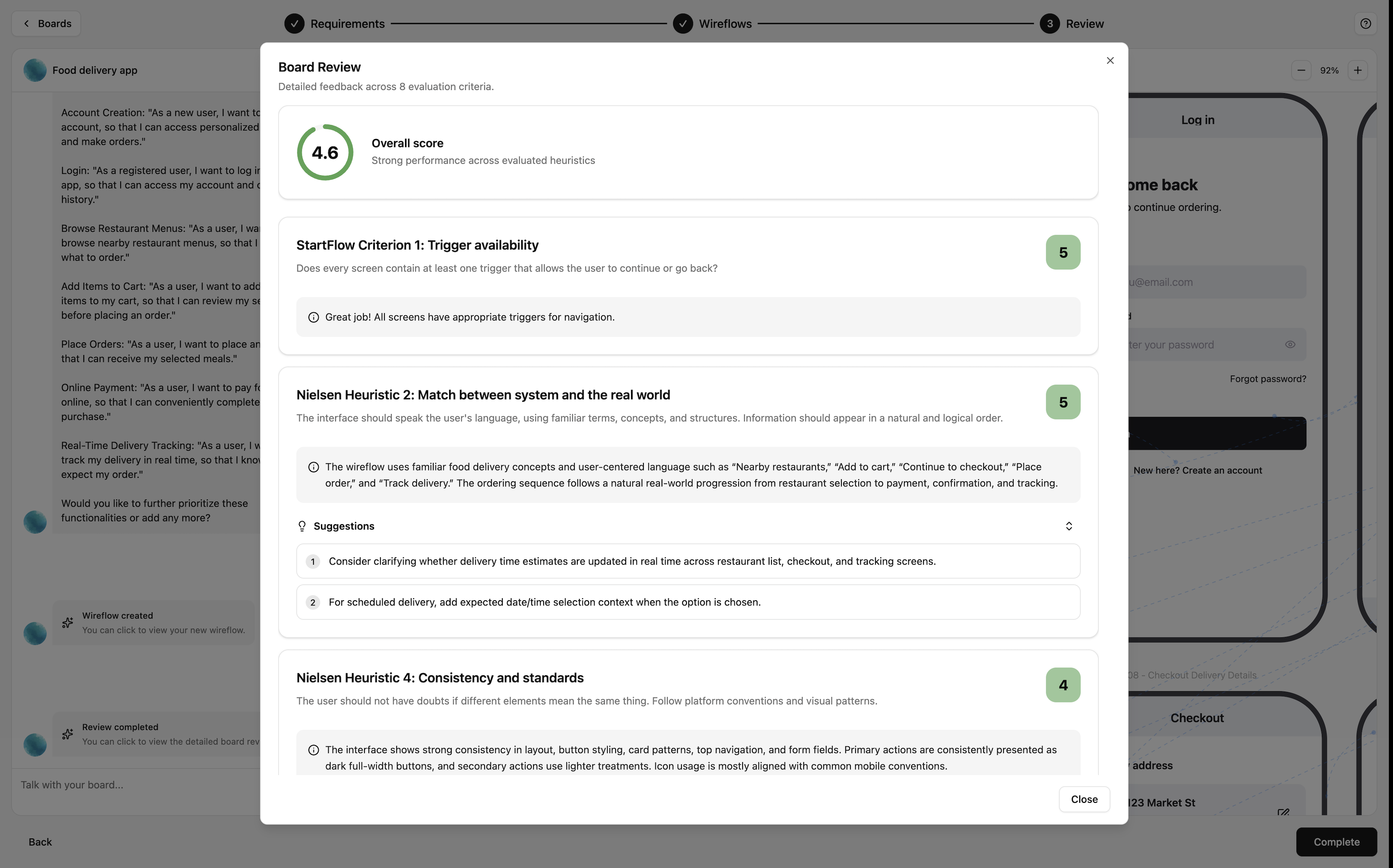}
\caption{Automatic UX review panel with scores and improvement suggestions.}
\Description{A modal popup titled "Board Review" showcasing an automated UX evaluation report. At the top left, a circular gauge displays an overall score of 4.6. Below, the modal lists detailed evaluation criteria based on StartFlow and Nielsen Heuristics, complete with green numeric scores, diagnostic feedback text, and expandable bullet points providing actionable improvement suggestions.}
\label{fig:review}
\end{figure}

After editing is completed, the user advances to the review stage, where the evaluation agent analyzes the wireflows according to StartFlow criteria and Nielsen’s Usability Heuristics. Figure~\ref{fig:review} presents the \textit{Board Review} panel, displaying the overall UX score and a detailed breakdown of the evaluated criteria, accompanied by suggestions for improvement.

The described scenario demonstrates how SoftBoard integrates the three stages of the StartFlow method into a continuous workflow: guided requirements structuring, wireflow generation and refinement, and automatic UX evaluation. This workflow highlights the tool's potential to reduce the cognitive burden of prototyping, standardize the application of the method, and support teams with limited UX expertise in developing MVPs that are better aligned with user needs.

Additionally, the tool allows users to generate a public link for viewing completed boards, facilitating stakeholder communication and the sharing of prototyping results.

\section{Conclusion and Future Work}
This paper presented SoftBoard, a web-based tool designed to operationalize the StartFlow method and support the creation and evaluation of low-fidelity prototypes in software startups. The motivation for this work stems from the observation that, although User Experience is widely recognized as a critical factor for the success of digital products, startup teams often face constraints related to time, resources, and expertise that hinder the systematic adoption of structured UX practices during the early stages of development.

The tool was conceived based on findings from a previously conducted Systematic Mapping Study, which revealed important gaps in the ecosystem of low-fidelity prototyping tools. The main findings included the predominance of general-purpose tools adapted for prototyping, the limited incorporation of explicit methodological guidance, and the still incipient use of Artificial Intelligence as continuous support throughout the design process.

SoftBoard was developed to address these gaps by integrating, within a single environment, the StartFlow method, a dedicated wireflow editor, and an LLM-based multi-agent system that supports the entire prototyping process. This integration transforms a method originally supported only by textual documentation into an interactive and guided workflow, reducing application variability and dependence on practitioners' individual experience.

The results presented indicate that the tool contributes in three main ways. First, it promotes the explicit incorporation of methodological guidance by structuring activities into sequential stages aligned with StartFlow. Second, it introduces Artificial Intelligence as continuous support throughout the process through agents that assist with requirements elicitation, automatic wireflow generation, and heuristic-based UX evaluation. Third, it seeks to lower the entry barrier for teams with limited UX expertise by providing a guided workflow that supports structured decision-making from the earliest stages of product development.

As future work, a feasibility study of the tool is currently underway, involving usability testing and interviews with software professionals in real startup contexts. The results of this study are expected to provide insights into practitioners' perceptions, identify opportunities for improvement, and offer initial evidence regarding the adoption potential of SoftBoard in the MVP development process.

\section*{Artifact Availability}

All artifacts, including the demonstration video and diagrams, are available at:
\url{https://doi.org/10.5281/zenodo.20369601}.

%% The next two lines define the bibliography style to be used, and
%% the bibliography file.
\bibliographystyle{ACM-Reference-Format}
\bibliography{sample-base}

\end{document}